# Similarity analysis of DNA sequences through local distribution of nucleotides in strategic neighborhood

*Probir Mondal, Department of Computer Science, P. R. Thakur Govt. College,*
*Pratyay Banerjee, Department of Physics, P. R. Thakur Govt. College,*
*Debranjan Pal, IIT, Madras*
*and*
*Krisnendu Basuli, Deparment of Computer Science, West Bengal State University*

*Abstract*—We propose a new alignment-free algorithm by constructing a compact vector representation on $\mathbb{R}^{24}$ of a DNA sequence of arbitrary length. Each component of this vector is obtained from a representative sequence, the elements of which are the values realized by a function $\Gamma$. This function $\Gamma$ acts on neighborhoods of arbitrary radius that are located at strategic positions within the DNA sequence and carries complete information about the local distribution of frequencies of the nucleotides as a consequence of the uniqueness of prime factorization of integer. The algorithm exhibits linear time complexity and turns out to consume significantly small memory. The two natural parameters characterizing the radius and location of the neighbourhoods are fixed by comparing the phylogenetic tree with the benchmark for full genome sequences of fish mtDNA datasets. Using these fitting parameters, the method is applied to analyze a number of genome sequences from benchmark and other standard datasets. Our algorithm proves to be computationally efficient compared to other well known algorithms when applied on simulated dataset.

## I. INTRODUCTION

Recent advancement in molecular biology stems primarily from the inclusion of Information Science and Technology into this field. So far, various computational tools have been developed and applied to analyze biological sequences. There exist algorithms to identify the origin of viruses, the quantum of similarity present among species, the mutation occurring inside them etc. The pioneering work in the field of sequence alignment was done by Needleman-Wunsch [1] in 1970 followed by Smith-Waterman [2]. Following them, alignment-based searching tools, viz., BLAST [3, 4], FASTA [5, 6] etc. were developed.

Molecular biology continued to become increasingly interesting with the enormous growth of its database generated by various initiatives. Despite their accuracy, most of the alignment-based algorithms that have been implemented so far admit quadratic time complexity and, thus, do not seem to have proved useful for analyzing long sequences. Consequently, it seemed customary to develop time-efficient algorithms. A new variety of alignment-free (AF) algorithms [7–13] arose thereafter as an alternative description of biological sequences with a focus to represent them through concrete mathematical object amenable to further treatment. AF algorithms, in general, admit linear time complexity. Moreover, their global nature is expected to play a vital role while comparing DNA sequences of unequal length.

A typical DNA sequence comprises of four nucleotide bases, viz., Adenine (A), Guanine (G), Cytosine (C) and Thymine (T). To compare two such sequences, there exist in the literature a large number of AF approaches that fall in two broad categories: viz., the word-based method [14, 15] which is based on the frequencies of subsequences of certain length and information-theory based method [16, 17] which quantifies the informational content between pair of sequences. In addition, there exist AF methods belonging to neither of these two categories. This includes, for example, techniques relying on the length of matching words viz., average common word [18], shortest unique substring [19], the minimal absent words between sequences [20], iterated maps [21], graphical representation [22, 23], chaos game representation [24]: all in the interest to extract information about the distribution of nucleotides within sequences.

Here, in this article, we adopt a new AF approach. Given a DNA sequence, instead of looking at individual nucleotide, we consider its corresponding course-grained form which provides a 'compact' representation of the same in the sense that the length of its corresponding representative sequence is half or even smaller compared to that of the sequence itself. This is a clear indication that the practical running time of our algorithm will decrease.

The arrangement of the paper is as follows: In Sec. 2, we propose our algorithm in detail regarding the association of a scalar with a DNA sequence of arbitrary length and calculation of the Euclidean metric ρ between a pair of such sequences. In Sec. 3, we apply our algorithm on full genome sequence of fish mtDNA to fix the values of the two fitting parameters. To do so, we calculate the normalized Robinson-Foulds (nRF) distance and the normalized Quartet Distance (nQD) between the phylogenetic tree obtained from our algorithm and that for the fish mtDNA taken from benchmark datasets [25]. In the next section, we explicitly demonstrate the efficiency of our method on five complete genome sequences by comparing running times with two well-known AF algorithms. Moreover, we compare graphically the running time and peak memory consumed by our method with the same two algorithms on simulated datasets. In Sec. 5, we finally draw conclusion and state possible future direction.

## II. CONSTRUCTION OF THE REPRESENTATIVE VECTOR

Consider a typical string ξ consisting of N nucleotides. We denote such a string $\xi = s_1 s_2 \cdots s_N$, where each $s_i$ is one of the four nucleotides, viz., A, C, G and T. Let P be the ordered set containing the first four prime numbers, i.e., P= {2, 3, 5, 7} and let $\sigma_i$ be the i[th] permutation of the set P. Clearly, we have 4! = 24 such permutations, viz., $\sigma_0, \sigma_1, \ldots, \sigma_{23}$; where we have chosen $\sigma_0$ to be the identity permutation. Corresponding to a particular permutation $\sigma_i$ of the set P, we assign to each of the nucleotides, the first four prime numbers as follows: $A = \sigma_i(1)$, $C = \sigma_i(2)$, $G = \sigma_i(3)$ and $T = \sigma_i(4)$, where $\sigma_i(j)$ is the j[th] element corresponding to the i[th] permutation of the set P. For example, in the case of identity permutation, we have i = 0 and thus $A = \sigma_0(1) = 2$, $C = \sigma_0(2) = 3$, $G = \sigma_0(3) = 5$, $T = \sigma_0(4) = 7$. Next, we define a l-neighbourhood $U_l(s_i)$ of the nucleotide $s_i$ with radius $l$ and centre $s_i$ to be the set $U_l(s_i) \equiv \{s_{i-l}, \ldots, s_i, \ldots, s_{i+l}\}$. Let $\Gamma_j$ be the function associating with each neighbourhood $U_l(s_i)$, a positive integer $\Gamma_j(U_l(s_i))$ with respect to the permutation $\sigma_j$ through the following way:

$$\Gamma_j(U_{l(s_i)}) = [\sigma_j(1)]^{f_1} \cdot [\sigma_j(2)]^{f_2} \cdot [\sigma_j(3)]^{f_3} \cdot [\sigma_j(4)]^{f_4} \quad (1)$$

where the exponents exponents $f_1$, $f_2$, $f_3$ and are the frequencies of the four nucleotides A, C, G and T, respectively in the neighbourhood $U_l(s_i)$. In other words, the value of Γ at a neighbourhood is simply the product of the prime numbers (PPN) assigned (according to some permutation) to the nucleotides appearing in that neighbourhood. Note there is no nucleotide $s_i$ for $i < 1$ or $i > N$. Clearly, $0 \leq f_r \leq 2l + 1$ for $r \in \{1, 2, 3, 4\}$. From the uniqueness of the prime factorization of a positive integer, it follows that the number $\Gamma_j(U_l(s_i))$ contains complete information about the frequencies of the nucleotides in the neighbourhood $U_l(s_i)$.

However, it is observed that given the number $\Gamma_j(U_l(s_i))$, one can reconstruct the associated neighbourhood $U_l(s_i)$ with the exact ordering of nucleotides only in case the degeneracy in the ordering of nucleotides is lifted depending upon the two values of Γ at the two overlapping neighbourhoods on either side of $U_l(s_i)$. This point is illustrated at the end of the present section.

Let
$$\zeta = \{\Gamma_j(U_l(s_1)), \Gamma_j(U_l(s_{t+2})), \Gamma_j(U_l(s_{2t+3})), \ldots, \Gamma_j(U_l(s_w))\} \quad (2)$$

be a sequence consisting of $n$ numbers $\Gamma_j(U_l(s_i))$, where
$$n = 1 + \lfloor \tfrac{N-1}{t+1} \rfloor \quad (3)$$

and $w = (n-1)t + n$. Here $\lfloor z \rfloor$ denotes the integer part of z and t is the distance (i.e., number of nucleotides) between the centres of two successive neighbourhoods. In Sec. 3, for a given value of l, we shall set $1 \leq t \leq l$. Otherwise, two adjacent neighbourhoods will no longer overlap and may lead to loss of information. We claim the sequence ζ in Eq. (2) to be a compact representation of the string ξ since from Eq. (3) we find n ≲ N/2. Next, our objective is to associate a scalar with the sequence (2) in a way such that the scalar is sufficiently sensitive as the string ξ undergoes point mutation, insertion and deletion. To this end, we propose the scalar $\eta_j$ (with respect to the permutation $\sigma_j$) associated with the sequence ζ by adding all the entries of ζ as follows:

$$\eta_j = \sum_{i=1}^{n} \Gamma_j\left(U_l(s_{i+t(i-1)})\right) \quad (4)$$

In order not to give preference to any particular permutation we assign prime numbers to the nucleotides A, C, G and T corresponding to every permutation $\sigma_j$ of the set P. Thus, for every $\sigma_j$ we obtain a scalar $\eta_j$ from Eq. (4). In this way a representative vector $\vec{\eta} = (\eta_0, \eta_1, \ldots, \eta_{23})$ is constructed corresponding to the string ξ. As each $\eta_j$ is real, we presume the vector $\vec{\eta}$ resides in the 24-dimensional real Euclidean space $\mathbb{R}^{24}$ endowed with the Euclidean metric ρ. Thus two DNA sequences are compared by computing ρ between the corresponding representative vectors.

As an example, let us consider a typical string of nucleotides

$$ACTGCCTCGATAA \quad (5)$$

Here N=13. Choose, say, the identity permutation $\sigma_0$. Then it follows that A=2, C=3, G=5 and T=7. Set the two parameters as $l$=1 and t=1. This means that we are considering neighbourhoods with centre at every alternate nucleotide comprising only of the nearest neighbours. The neighbourhoods for the string (5) (see Fig. 1) are as follows $U_1(s_1) = \{A, C\}$, $U_1(s_3) = \{C, T, G\}$, $U_1(s_5) = \{G, C, C\}$, $U_1(s_7) = \{C, T, C\}$, $U_1(s_9) = \{C, G, A\}$, $U_1(s_{11}) = \{A, T, A\}$ and $U_1(s_{13}) = \{A, A\}$.

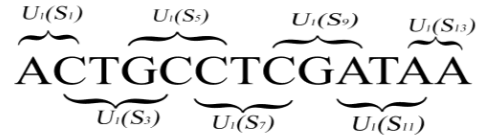

Fig. 1 A typical string of 13 nucleotides. The collection of nucleotides lying inside an over/under brace forms a neighborhood with centre at the middle nucleotide and radius $l = 1$. However, the two extreme nucleotides are the centres of the neighborhoods $U_1(s_1)$ and $U_1(s_{13})$, respectively as $U_1(s_1)$ has no nucleotide on the left of its centre at $s_1$ and $U_1(s_{13})$ has no nucleotide on the right of $s_{13}$. The distance $t$, i.e., the number of nucleotides between successive neighbourhoods, is 1.

Note that the neighbourhoods at the two extreme positions naturally contain fewer nucleotides. Using Eq. (1), we calculate, say, the number $\Gamma_0(U_1(s_5))$ with respect to the identity permutation $\sigma_0$ as $\Gamma_0(\{G, C, C\}) = 2^0 \cdot 3^2 \cdot 5^1 \cdot 7^0 = 45$. In this way, as shown in Eq. (2), we construct the sequence $\zeta$ as

$$\zeta = \{6, 105, 45, 63, 30, 28, 4\}. \tag{6}$$

Using Eq. (3), we find n = 7. Finally, from Eq. (4), we find the scalar $\eta_0$ (associated with the string (5) and which is simply the sum of the elements of the sequence (6)) is 281. As stated earlier, we call the sequence (2) to be a compact representation of the string $\xi$ as the former contains around half or even smaller number of elements compared to the latter. In addition, the uniqueness of prime factorisation of integer reproduces the local distribution of the nucleotides with exact frequencies. However, the invertibility of this representation depends heavily on the entries of the sequence (2). As an illustration, choose a subsequence $\zeta' = \{6, 105\}$ of the sequence $\zeta$ in (6). As $6 = 2^1 \cdot 3^1 \cdot 5^0 \cdot 7^0$, the corresponding neighbourhood, which is the inverse image of the function $\Gamma$, is either $\{A, C\}$ or $\{C, A\}$. Similarly for the second element of $\zeta'$, we find $105 = 2^0 \cdot 3^1 \cdot 5^1 \cdot 7^1$, so that the corresponding neighbourhood is one of the 6 permutations of C, G and T, viz., $\{C, G, T\}$, $\{C, T, G\}$, $\{G, C, T\}$, $\{T, C, G\}$, $\{G, T, C\}$ and $\{T, G, C\}$. Note that the nucleotide A is absent in all of the six neighbourhoods. Thus, looking at the overlap region, it is clear that $\{A, C\}$ is the right choice of neighbourhood corresponding to the element 6 of $\zeta'$. Consequently, $\{C, G, T\}$ and $\{C, T, G\}$ are the remaining possibilities as the inverse image of $\Gamma$ of the second element 105 of $\zeta'$. Thus, from the subsequence $\zeta'$, we reproduce (up to a degeneracy in the ordering of the last two nucleotides) the corresponding string of nucleotides to be either ACGT or ACTG. Notice that the example (5) that we have chosen here is perfectly reproducible following this procedure from its representative sequence (6).

## III. SIMILARITY ANALYSIS

In order to test the usefulness of our algorithm, proposed in Sec. 2 and denoted as PPN hereafter, we apply it on the benchmark dataset of fish mtDNA from AFproject [25]. The corresponding phylogenetic tree obtained through PPN is shown in Fig. 2 by setting the parameters at $l = 4$ and $t = 1$. Upon comparison with the benchmark, in the case of fish mtDNA, we find for PPN, the nRF is 0.64 and nQD is 0.2602. Implementing the Manhattan distance (in place of Euclidean distance) modifies the nQD value only slightly. In this case it is 0.2723. It is to be noted that PPN is expected to operate upto $l = 4$ as the fluctuation in the distribution of frequencies of different nucleotides in a neighborhood becomes smaller with increasing radius.

Choosing $l = 4$, $t = 1$, we apply PPN on the assembled 29 E. Coli/Shigella strain taken from the Genome-based phylogeny (GBP) section and on Yersinia, plants and unsimulated 27 E. Coli/Shigella strain from the Horizontal Gene Transfer (HGT) section of the benchmark datasets.

In addition, using the datasets from ref. [26], we generate the phylogenetic tree shown in Fig. 3 through PPN for the mammals mtDNA. Implementation of our algorithm is available at https://github.com/Workspace-PM/PPN.

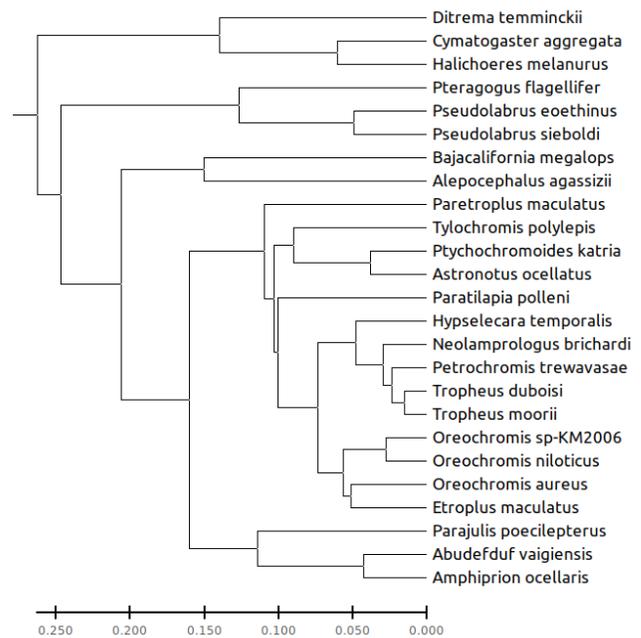

Fig. 2 UPGMA tree for fish mtDNA sequences having 25 species drawn in software MEGA 11 with parameter values $l = 4$, $t = 1$.

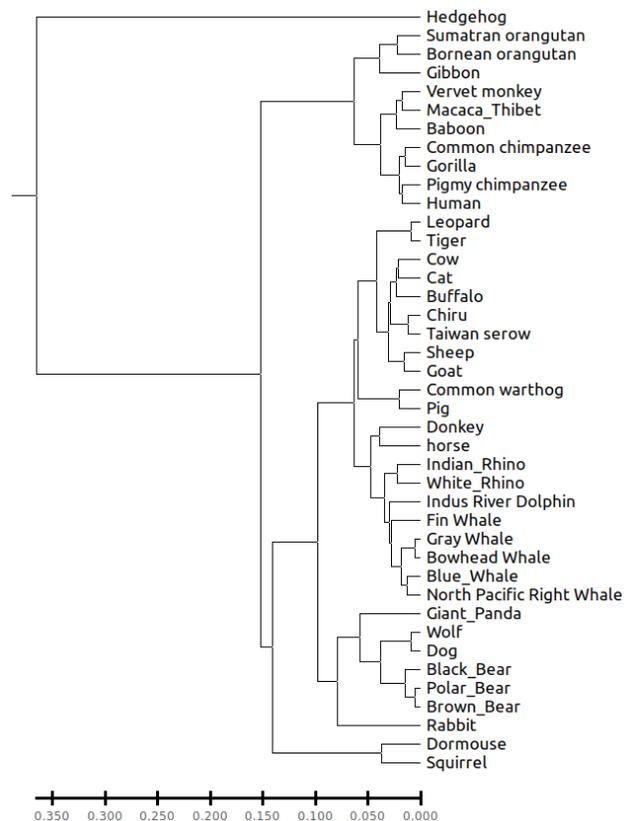

Fig. 3 UPGMA tree for mammals mtDNA sequences having 41 species drawn in software MEGA 11 with parameter values $l = 4$, $t = 1$.

## IV. PERFORMANCE ANALYSIS

In this section we demonstrate the efficiency of PPN proposed in Sec. 2 by comparing the running time against standard AF tools. We have conducted the computation on a machine with Intel(R) Core (TM) i7-8700 CPU @ 3.20 GHz x 12, 16 GB RAM and driven by Ubuntu 22.04.03 OS. The comparison of the runtimes taken by PPN, CDMAWS [13] and Co-phylog [10] for complete genome sequences of mammals, influenza A virus, rhinovirus, ebolavirus and coronavirus are shown in the Table 1. It is apparent from this comparison that PPN, which admits a linear time-complexity, is, indeed, fast. Note the value of the parameter t provides an estimate of the extent to which one can compactify the representation of a given DNA sequence. For example, when t is set to be 1, i.e., at its minimum value, we find the length n in Eq. (3) of the sequence $\zeta$ becomes approximately half that of the original DNA sequence $\xi$.

Additionally, we run PPN on simulated datasets generated by Seq-gen Monte Carlo simulation tool [27] where each sequence representing a particular species contains 50,000 nucleotides. In Fig. 4 we plot the variation of the average running time taken by three algorithms (including PPN) with the number of species. The average running time is calculated by taking the average of 10 runs conducted on a machine with Intel(R) Core (TM) i7-8700 CPU @ 3.20 GHz x 12, 16 GB RAM and driven by Ubuntu 22.04.03 OS. The number of species is varied over a wide range from 100 to 900. Similarly, in Fig. 5, the peak memory consumption is recorded by choosing the maximum value from 10 trials. On this simulated dataset, PPN seems to be computationally efficient both in terms of runtime and peak memory consumption compared to Co-phylog and CD-MAWS within the range shown in Fig. 4 and Fig. 5. It is to be noted that the combinatoric factor representing the number of actual pairs of sequences to be compared within a dataset increases quadratically with the number of species.

Moreover, to unravel the truly alignment free nature of PPN, we apply it to compare two sequences differing drastically in the number of nucleotides. One chromosome of Zea mays (GCF 000005005.2) contains 30,70,41,717 nucleotides while that of Oryza sativa (GCF 001433935.1) contains 4,32,70,923 nucleotides. PPN runs successfully and takes 33.68 minute to determine the distance between these sequences and consumes maximum 20.24 GB memory with Intel(R) Xeon(R) GOLD 6134 CPU @3.20GHz x 16, 64GB RAM and Ubuntu 22.04 OS.

TABLE I
RUNNING TIME (IN MINUTES) COMPARISON OF FIVE COMPLETE GENOME SEQUENCES

| Method | Mammals | Influenza | Rhino | Ebola | Corona |
|---|---|---|---|---|---|
| PPN | 0.052 | 0.004 | 0.067 | 0.086 | 0.071 |
| CD-MAWS | 0.061 | 0.013 | 0.162 | 0.058 | 0.057 |
| Co-phylog | 0.151 | 0.116 | 0.283 | 0.216 | 0.125 |

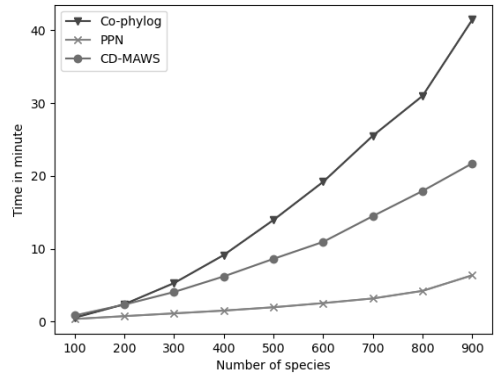

Fig. 4 Average running time (in minute) of PPN for simulated datasets with number of species ranging from 100 to 900. Each species contains 50,000 nucleotides.

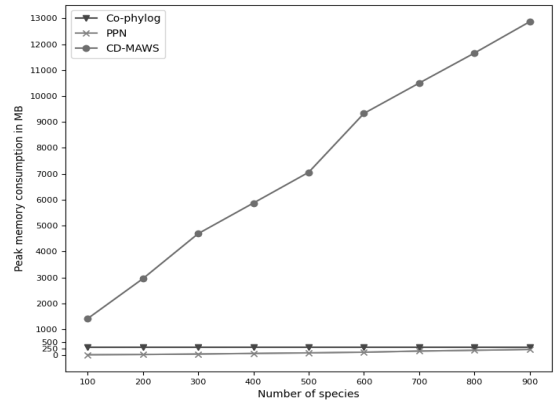

Fig. 5 Peak memory consumption in MB by PPN for simulated datasets with number of species varying between 100 and 900. Here, each species contains 50,000 nucleotides.

## V. CONCLUSION

In this article, we have adopted a new alignment-free approach to trace the amount of similarity/dissimilarity present within a pair of DNA sequences. PPN associates a scalar in a simple way to the representative sequence $\zeta$ corresponding to the string $\xi$. The problem regarding comparing sequences of unequal length is circumvented through the construction of the said scalar. The significant reduction in running time has occurred in two steps. As a first step, we have considered the course-grained form of the original DNA sequence $\xi$ to reduce its length to a considerable extent. Further reduction happens due to the asymptotically linear nature of the time complexity arising in our algorithm. Comparison with standard AF algorithms clearly reveals that PPN is fast and its peak memory consumption is significantly small when applied on both real and simulated datasets. Due to the minimal memory requirement, the method is believed to be implemented by a larger group of researchers. Moreover, PPN can be used to extract features from sequences which is essential for the construction of models using machine learning techniques.